\documentstyle[twocolumn,prb, epsf, aps]{revtex}

\begin{document}
\draft

\twocolumn[\hsize\textwidth\columnwidth\hsize\csname @twocolumnfalse\endcsname

\title{Experimental evidence of the spin selection rule in KLL Auger transition}
\author{D. J. Huang,$^{1}$ W. P. Wu,$^{2}$  C. F. Chang,$^{1}$ J. Chen,$^{3}$ S. C. Chung,$^{1}$\\
 L. H. Tjeng,$^{4}$ and C. T. Chen$^{1}$}
\address{$^{1}$Synchrotron Radiation Research Center, Hsinchu 30077, Taiwan\\
$^{2}$Dept. of Physics, National Chang Hua University of Education 500, Chang-Hua, Taiwan\\
$^{3}$Department of Physics, National Chung Cheng University, Chia-Yi 621, Taiwan\\
$^{4}$Solid State Physics Lab., Materials Science Center, University of Groningen,\\ Nijenborgh 4, 9747 AG Groningen, The Netherlands }

\author{S. G. Shyu,$^{5,6}$ and C. C. Wu,$^{5}$} 
\address{$^{5}$Institute of Chemistry, Academia Sinica, Taipei 115, Taiwan\\
$^{6}$Department of Chemistry, National Central University, Tao-Yuan 320, Taiwan\\ }
\date{\today }
\maketitle
\begin{abstract}
With on- and off- resonant excitation photons, spin-resolved Auger electron spectra of epitaxial CrO$_{2}$ thin films
show an experimental evidence of the 
spin-selective KLL Auger decay. The on-resonance O KLL Auger electrons are found 
to be highly spin-polarized, while the off-resonance ones with almost zero spin polarization. 
These results lead to conclude that the two-hole final state in KLL Auger decay
is a spin-singlet. Applications to spin-resolved absorption spectroscopy are discussed.
\end{abstract}

] \narrowtext 

Spin-resolved core-level spectroscopy has been shown very effective to
unravel the spin-dependent electronic structure of materials.\cite{Johnson97}
For instance, the Zhang-Rice spin-singlet in high temperature
superconductors (HTSC) has been experimentally confirmed using spin-resolved
resonant photoemission techniques.\cite{Tjeng} Magnetic circular dichroism
(MCD) in x-ray absorption\cite{Erskine75,Chen} and spin-resolved Auger
spectroscopies are another two examples.\cite
{Landolt82,Landolt85,Taborelli86,Allenspach87} These two techniques have
drawn much attention to the exploring local element-specific magnetic
properties of materials. For instance, MCD has been widely used because of
its unique feature of obtaining the information on element-specific magnetic
spin and orbital moments.\cite{Thole92,Cara93,Chen95}

Being a finger-print technique, Auger electron spectroscopy is very powerful
and widely used for chemical analysis of thin films and surfaces.\cite
{Fuggle81} With the analysis of the spin polarization of Auger electrons
from magnetic materials, the traditional Auger spectroscopy can be extended
to spin-polarized Auger spectroscopy for exploring magnetic thin films and
surfaces. It has been demonstrated that the spin polarization of
core-valence-valence (CVV) Auger electrons is a direct manifestation of the
local spin polarization of the valence electrons.\cite{Landolt82} Auger
transition contains valuable information on the electron correlations and
the core-hole screening by valence electrons. It has been shown that an
important spin-dependent interaction, or exchange correlation, exists in the
final state of the LMM Auger transition in $3d$ itinerant ferromagnets such
as Fe. The exchange correlation in LMM Auger transition results in that the
probability the spins of the two holes in the Auger final state being
parallel is comparable to that of being antiparallel.\cite{Sinkovic95} On
the other hand, several theoretical works have shown that the spin triplet
in the KLL Auger transition is forbidden, and only the spin-singlet is
allowed.\cite{Fuggle81,Sawatzky} To our knowledge, at present, no direct
experimental results are available to show this spin selection rule of KLL
Auger transition.

In this paper, we report a direct experimental evidence of the exchange
correlations in KLL Auger decay With on- and off- the O K-edge-resonant (1s$%
\rightarrow $2p) excitation, the dramatic difference in the spin
polarization shows that the exchange correlation effect prevents the
two-hole Auger final state from being a spin-triplet. This exchange
correlation effect in KLL Auger decay leads to a very important application
to spin-resolved x-ray absorption of magnetic oxides in which oxygen plays
an important role in magnetism via superexchange or double exchange
interactions.

The spin resolved Auger spectroscopy measurements were carried out at the
elliptically polarized undulator (EPU)\cite{Hwang99} beamline of the
Synchrotron Radiation Research Center (SRRC), Taiwan. The light polarization
can be adjusted to be circularly polarized or linearly polarized with the
electric vector of the light either in the horizontal or the vertical plane,
with taking the advantage of the EPU.\cite{Chung01} The spin-resolved Auger
electrons were collected at the normal emission of epitaxial CrO$_{2}$ films
at the remnant state at temperature of 80${\rm K}$. A commercial
hemispherical energy analyzer was used to energetically analyze Auger
electrons; a micro-Mott spin detector adapted to the energy analyzer was
used for spin analysis. The instrumental asymmetry of the spin-detector was
eliminated by reversing the magnetization of the sample.

Epitaxial CrO$_{2}$ films were grown on TiO$_{2}$(100) substrates using the
chemical vapor deposition (CVD) technique. The substrates were
ultrasonically cleaned with acetone, 1,1,1-trichloroethane, 10\%
hydrofluoric acid and distilled water subsequently before air-drying. The
deposition was performed at 400$^{\circ }{\rm C}$ by using CrO$_{3}$ as the
precursor.\cite{Suzuki99,Li99} The film thickness was estimated to be around
2000\AA . The CrO$_{2}$ films were characterized as epitaxy by x-ray
diffraction. Fig.1(a) displays the normal $\theta -2\theta $ scan in the
vicinity of the (200) peak of the film and the substrate. The FWHM of the
rocking curve of the CrO$_{2}$ (200) peak is 0.2$^{\circ }$, showing good
quality of the crystalline thin films. The magnetic properties were examined
using magnetic optical Kerr effect (MOKE) measurements of the films. As
shown in Fig. 1(b), the MOKE measurements indicate that the magnetic easy
axis of the films is along [001] direction, that the magnetic properties of
CrO$_{2}$ films are the same as those of bulk CrO$_{2}$, consistent with the
results of previous epitaxial CrO$_{2}$ thin films.\cite{Li99} Ex-situ soft
x-ray polarization-dependent absorption measurements of CrO$_{2}$ samples
were performed in an ultrahigh vacuum (UHV) chamber without further
treatments of annealing or sputtering.
\begin{figure}[tbp]
\centerline{\epsfxsize=3.0in \epsfbox{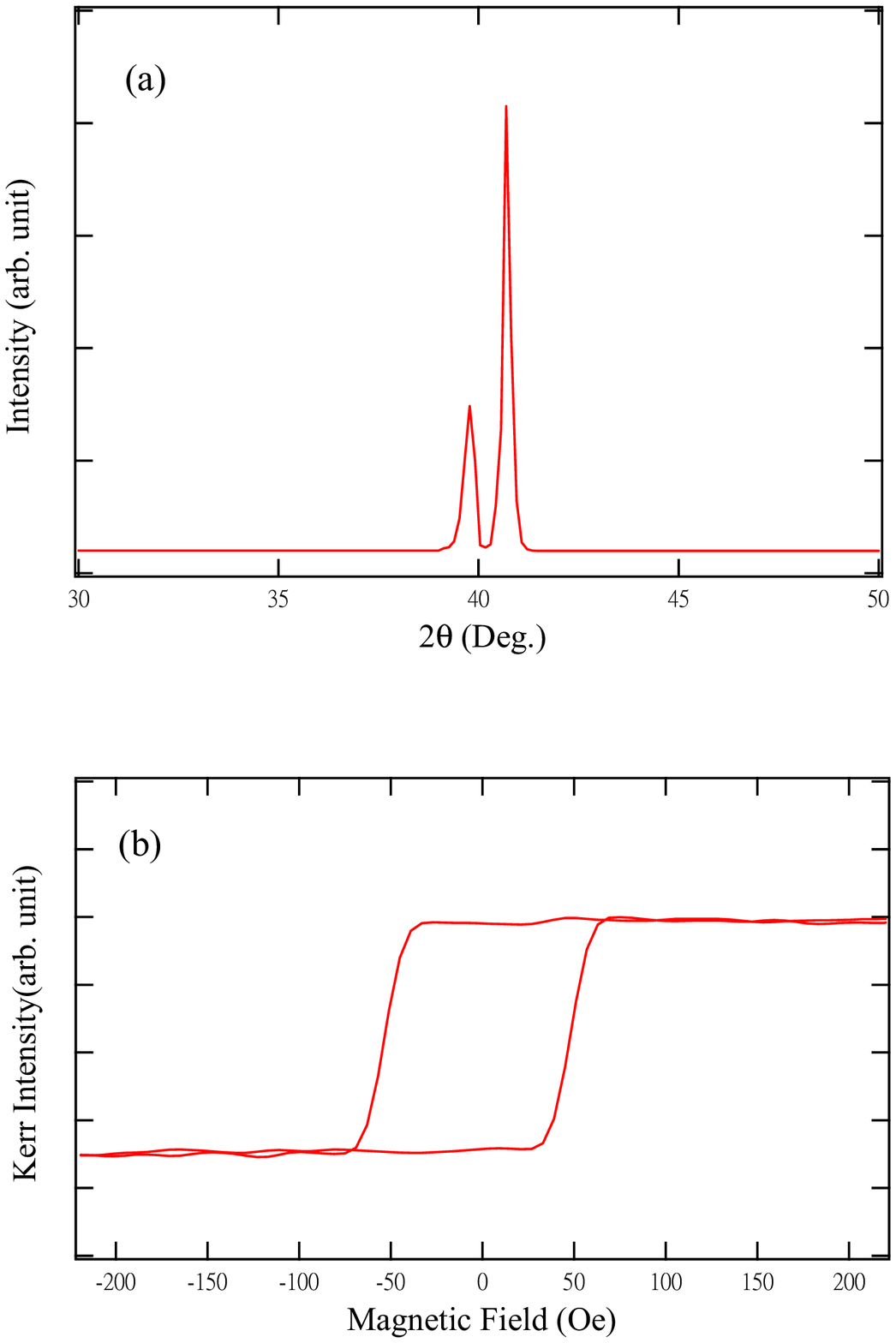}}
\caption{(a) Normal $\protect\theta-2\protect\theta $ x-ray diffraction
curve of CrO$_{2}$ thin films epitaxially grown on TiO$_{2}$ (100)and (b) MOKE measurements of CrO$_{2}$ thin films with the applying
magnetic field along [001].}
\end{figure}

Recent first-principles calculations\cite{Lewis97,Korotin98,Mazin99} and
contact measurements\cite{Soulen98,Ji01} have concluded that CrO$_{2}$ is a
half-metallic ferromagnet in which one spin channel is conductive, but the
other spin channel is insulating. The unoccupied states of CrO$_{2}$,
therefore, is fully spin-polarized. Recent LSAD+U calculations\cite
{Korotin98} has concluded that there is a strong hybridization between O 
{\it 2p} and Cr {\it 3d}. Fig. 2(a) depicts the O {\it 1s} absorption of CrO$%
_{2}$ with the {\bf E} vector of the light perpendicular to the magnetic
easy axis. The exceptionally large intensity in the O {\it 1s} absorption 
\cite{Stagarescu00}leads to the experimental confirmation of the strong
hybridization between O {\it 2p} and Cr {\it 3d}.\cite{} The de-excitation
of the O 1{\it s} absorption final state is dominated by an Auger decay,
leaving a two-hole final state. The Auger electron intensity is, therefore,
a manifestation of the O {\it 1s} absorption cross section.
\begin{figure}[tbp]
\vspace{5mm}
\centerline{\epsfxsize=3.0in \epsffile{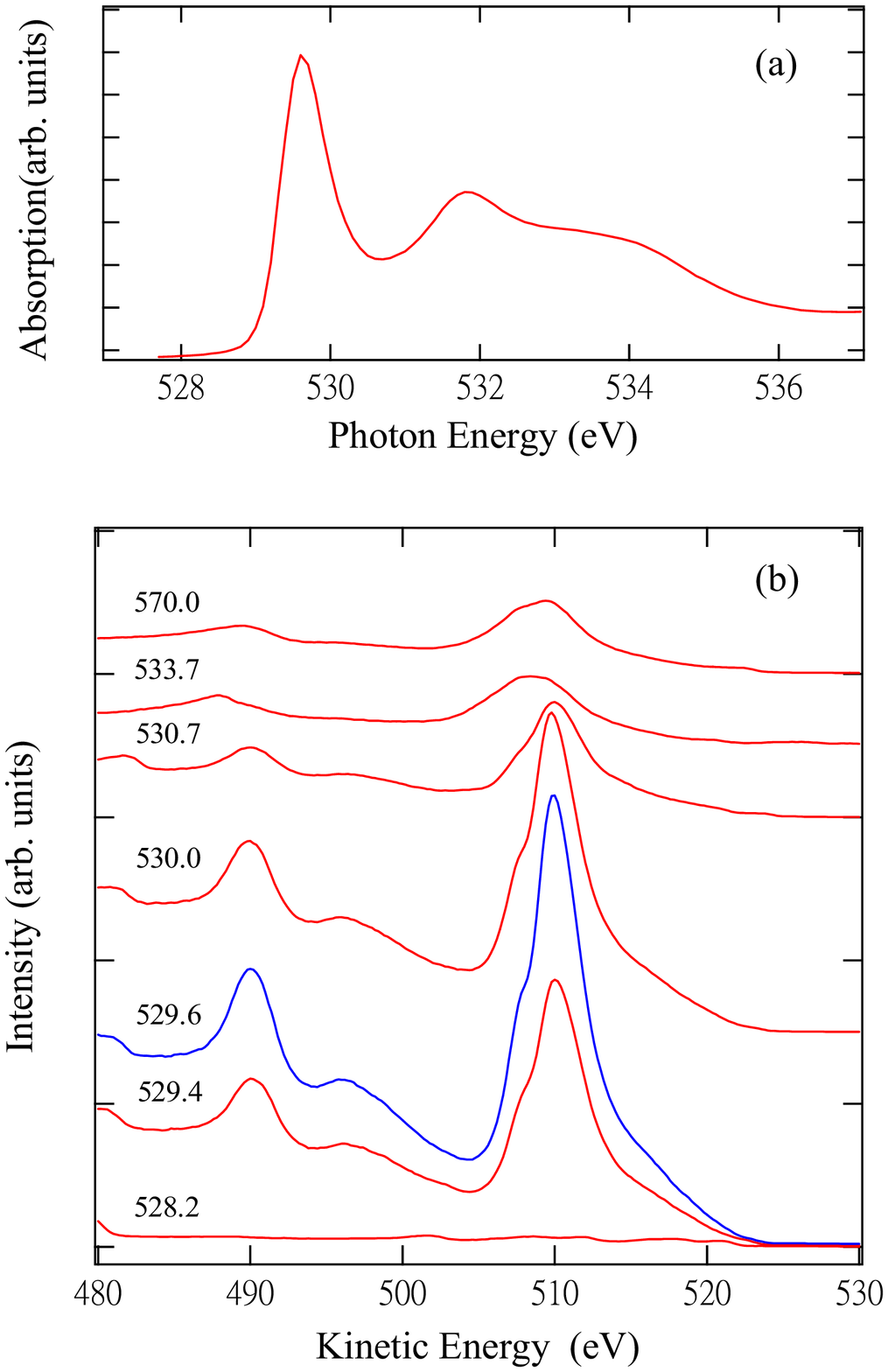}}
\caption{(a) O {\it 1s} absorption spectrum of CrO$_{2}$ thin films and (b)
a series of Auger spectra as the excitation photon energy around the
absorption peak, and 40.6 eV above. The photon energy for each Auger spectrum is shown in unit of eV.}
\end{figure}

Fig 2(b) shows a series of Auger spectra excited by photons with energies
around the O {\it 1s} absorption peak, and 40.6 eV above the peak energy.
The Auger electron intensity is resonant at h$\upsilon $=529.6 eV where the
Auger intensity is $\sim $ 50 times of the valence intensity. With the
excitation photons tuned on the resonance, the spin polarization of the O 
{\it 1s} core hole resembles that of the O {\it 2p} hole since spin is
conserved in a photo-excitation process. The spin polarization of the Auger
electrons depends on the spin of the O {\it 1s} hole and on the exchange
correlation in the two-hole Auger final state. For the spin-triplet Auger
two-hole final state, the spin of Auger electron is parallel to that of the 
{\it 1s} core hole. For the spin-singlet Auger two-hole final state, the
spin of Auger electron, on the other hand, is antiparallel to that of the 
{\it 1s} core hole.

The cross section of KLL Auger transition $\sigma _{_{Auger}}$ can be
approximately written as 
\begin{equation}
\sigma _{_{Auger}}\approx \left| \left\langle e^{i{\bf k}\cdot {\bf r}_{{\bf %
1}}}\cdot 1\left| \frac{1}{\left| {\bf r}_{1}-{\bf r}_{2}\right| }\right|
p_{_{1}}p_{_{2}}\right\rangle \right| ^{2},
\end{equation}
where {\it p}$_{1}$ and {\it p}$_{2}$ are the two {\it 2p} states involved
in the Auger decay, {\bf r}$_{1,2}$ and {\bf k} are the position vectors and
the wave vector of the Auger electrons, respectively. As a result of the
Auger decay, the electron at {\bf r}$_{1}$ is emitted to be the Auger
electron, and that at {\bf r}$_{2}$ falls into the {\it 1s} state. When both
of the two {\it 2p} states are at the same orbital, say {\it 2p}$_{x1}$ and 
{\it 2p}$_{x2}$, the initial state of the Auger decay can be written as $%
\left| (x_{_{1}}x_{_{2}})\chi _{s}\right\rangle ,$ in\ which \ $\chi _{s}$\
stands for the spin singlet. For the {\it 2p} states at different orbitals,
say {\it 2p}$_{x1}$ and {\it 2p}$_{y2}$, the Auger initial states are $%
\left| (x_{_{1}}y_{_{2}}-x_{_{2}}y_{_{1}})\chi _{t}\right\rangle ,$ and $%
\left| (x_{_{1}}y_{_{2}}+x_{_{2}}y_{_{1}})\chi _{s}\right\rangle ,$ for the
spin triplet $\chi _{t}$ and the spin singlet $\chi _{s}$, respectively.
Taking into account the spin part and the symmetry requirement, the matrix
element for the spin triplet state can expressed as: 
\begin{equation}
\left\langle (e^{i{\bf k}\cdot {\bf r}_{{\bf 1}}}-e^{i{\bf k}\cdot {\bf r}_{%
{\bf 2}}})\chi _{t}\left| \frac{1}{{\bf r}_{12}}\right|
(x_{_{1}}y_{_{2}}-x_{_{2}}y_{_{1}})\chi _{t}\right\rangle ,
\end{equation}
where {\bf r}$_{_{12}}$= $\left| {\bf r}_{1}-{\bf r}_{2}\right| $, and the
matrix element for the spin singlet state can be written as:

\begin{equation}
\left\langle (e^{i{\bf k}\cdot {\bf r}_{{\bf 1}}}+e^{i{\bf k}\cdot {\bf r}_{%
{\bf 2}}})\chi _{s}\left| \frac{1}{{\bf r}_{12}}\right|
(x_{_{1}}x_{_{2}})\chi _{s}\right\rangle ,
\end{equation}
and 
\begin{equation}
\left\langle (e^{i{\bf k}\cdot {\bf r}_{{\bf 1}}}+e^{i{\bf k}\cdot {\bf r}_{%
{\bf 2}}})\chi _{s}\left| \frac{1}{{\bf r}_{12}}\right|
(x_{_{1}}y_{_{2}}+x_{_{2}}y_{_{1}})\chi _{s}\right\rangle ,
\end{equation}
when the Auger transition occurs at the same and the different orbitals,
respectively. It is straight forward to prove that the matrix element of the
triplet state vanishes as a result of the rational and the inversion
symmetry of $\frac{1}{{\bf r}_{12}}$. Therefore only the spin singlets are
allowed in KLL Auger decay, consistent with the results of previous
theoretical works.

Since the spin triplet of the two-hole final state in KLL Auger transition
is forbidden, the intensity of the spin-up and spin-down Auger electrons
with energy {\it E} can be approximately expressed by a convolution of the
spin-resolved density of state (DOS): 
\begin{equation}
I_{\uparrow (\downarrow )}=M_{\uparrow \downarrow }\int \int D_{\uparrow
(\downarrow )}(\epsilon ^{\prime })D_{\downarrow (\uparrow )}(\epsilon
)\delta (E-\epsilon -\epsilon ^{\prime }-U)d\epsilon d\epsilon ^{\prime },
\end{equation}
where $M_{\uparrow \downarrow }$ and $U$ are the Auger matrix element for
the spin singlet of the two-hole Auger final sate and the Coulomb
interaction energy between the two holes, respectively, $D_{\uparrow
(\downarrow )}(\epsilon ^{\prime })$ and $D_{\uparrow (\downarrow
)}(\epsilon )$ refer to as the spin-up \ (spin-down) electrons emitted to be
Auger electrons and those falling into the initial core hole, respectively.
On the resonance excitation of a material with 100\% spin-polarized
unoccupied states, the initial $1s$ core hole is completely spin-polarized,
and the KLL Auger electrons, therefore, are 100\% spin-polarized with the
sign opposite to that of the states above the Fermi level.

\centerline{\epsfxsize=3.0in \epsfbox{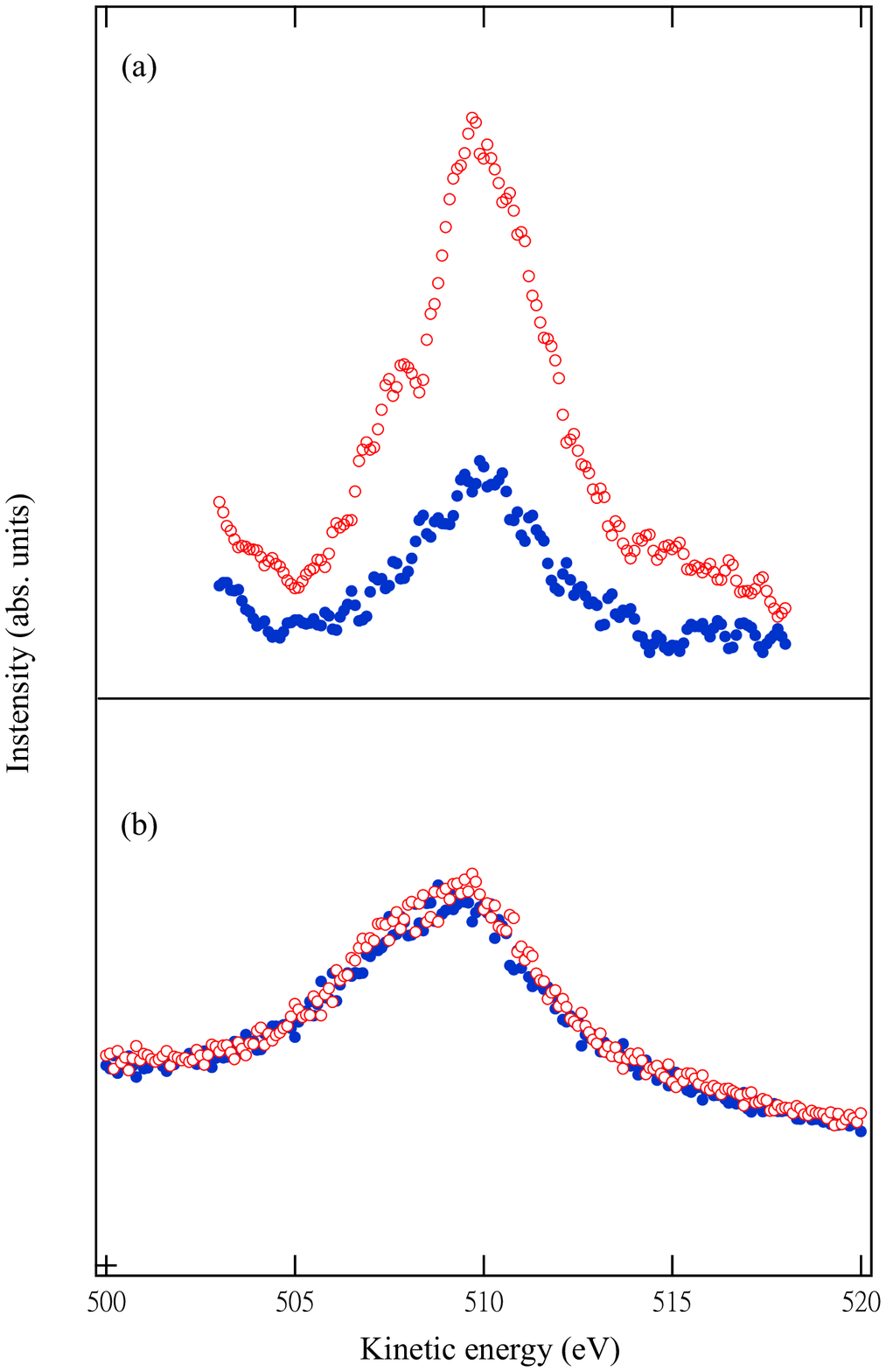}}
\begin{figure}[tbp]
\caption{Spin-resolved Auger spectroscopy measurements of CrO$_{2}$ thin
films. The Auger electrons are excited by (a) the photons being tuned
at the O 1s$\rightarrow 2p$ resonance, and by (b) the photon energy being tuned at 40.6 eV above the resonance. The filled and the open circles are for the spin-up and the spin-down results.}
\end{figure}
To verify the spin selection rule in KLL Auger decay discussed earlier, we
have performed on-and off-resonance spin-resolved Auger spectroscopy
measurements of CrO$_{2}$ thin films. As shown in Fig. 3(a), the Auger
electrons excited by the photons being tuned at the O 1s$\rightarrow 2p$
resonance are highly spin-polarized. Although the polarization of the lowest
major structure of the conduction band in a half-metallic material are
expected be 100\%, the spin polarization of the absorption of may be reduced
due to the electron correlation effects present in the system. In such a
system, the concept of density of states as defined in effective
one-particle theories loses its meaning as a basis for a quantitative
understanding of excitation spectra, i.e. density of states are then quite
different from spectral weights. It has been very recently shown that strong
electron-electron correlations set the maximal spin-polarization of the
lowest-energy O 1$s$ absorption peak of CrO$_{2}$ to be 50\%.\cite{Huang01}
In the following, we use a configuration-interaction cluster model to
briefly explain this calculation. On the resonance, the absorption is then
given by the transition from the ground state, which is a linear combination
of O($2p^{6}$)Cr($3d^{2}$) and O($2p^{5}$)Cr($3d^{3}$) states of $^{3}T_{1}$
symmetry, to the lowest XAS final state which is given by O($\underline{1s}%
2p^{6}$)Cr($3d^{3}$) of $^{4}A_{2}$ symmetry, where $\underline{1s}$ denotes
the core hole. With CrO$_{2}$ in the ferromagnetic state, the $3d^{2}$ ($%
^{3}T_{1}$) ground state has the quantum numbers $S=1$ and $S_{z}=1$. The $%
3d^{3}$ ($^{4}A_{2}$) electron addition state, however, can have the quantum
numbers $S=3/2$, $S_{z}=3/2$ and $S=3/2$, $S_{z}=1/2$, depending on whether
a spin-up or a spin-down electron is added, respectively. The fractional
parentage for the one-electron addition has a ratio of 3:1, so that the
spin-polarization for the $^{4}A_{2}$ peak is expected to be $\frac{3-1}{3+1}
$ = 50\%, which is very close to the measured value.

With the existence of the electron correlation effects in CrO$_{2}$, it is
crucial to assure the validity of the spin selection rule and that the
reduction in the spin polarization is indeed caused by the fractional
parentage effect. According to Eq. 5, the spin-up and the spin-down Auger
electrons excited by off-resonance photons are almost with equal intensity,
unlike the LMM Auger transition of a ferromagnet where both the spin-triplet
and the spin- singlet are allowed in the Auger final state. It has been
found that spin polarization of the off-resonance Fe LMM Auger electrons is
significant, \ $\approx $ 27\%. In contrast, Fig. 3(b) shows that the
spin-up and the spin-down KLL Auger electrons excited by the photon energy
being tuned at off-resonance, e.g. 570 eV, are equally populated, since the
initially created 1s hole is nearly non-spin-polarized. The zero spin
polarization of the O KLL Auger electrons supports that the spin-triplet in
KLL Auger transition is forbidden.

The spin-resolved Auger spectra with the excitation photons being tuned on
and off the resonance unambiguously show that the spin-triplet in KLL Auger
transition is forbidden. This spin selection rule provides us a powerful
means for spin-resolved O 1s absorption spectroscopy. With the spin
selection rule, the spin-resolved O 1s absorption spectroscopy can be
achieved by monitoring the spin resolved Auger electrons which are emitted
at a constant kinetic energy, i.e. the Constant-Final-State (CFS) mode. The
spin character of the O 2p states above the Fermi level can be revealed by
using spin-resolved O 1s XAS. This spin resolved technique is complementary
to the MCD in O K-edge absorption, which is predominately resulted from the
O 2{\it p}-projected orbital magnetization. The spin-resolved O $1s$
absorption spectroscopy is a very powerful technique to study magnetic
oxides in which the O 2{\it p} and the 3{\it d} of transition metals are
strongly hybridized, leading to their interesting magnetic and electrical
properties.\cite{Huang01,Steenenken}

In conclusion, the two-hole final state in KLL Auger transition has been
shown to be a spin-singlet as a consequence of the parity symmetry. The
results of spin-resolved Auger electron spectroscopy on half-metallic CrO$%
_{2}$ thin films with the photon energy being tuned on and off the O 1s$%
\rightarrow $2p resonance have revealed that the on-resonance O KLL Auger
electrons are highly spin polarized and the off-resonance ones are with spin
polarization close to zero. These results lead to conclude that the spin
triplet state $^{3}${\it P} of the two-hole state in KLL Auger decay is
forbidden, while the spin singlet states $^{1}${\it S} and $^{1}${\it D} are
allowed. With this spin selection rule, the spin-resolved of XAS in partial
Auger electron yield mode can be achieved. The spin-resolved XAS will be an
effective technique to unravel the spin feature of the unoccupied electronic
states in magnetic oxides.

The authors would like to acknowledge G. A. Sawatzky for sharing his
unpublished lecture note. We also would like to thank the SRRC staff for
their technical help. C.C. Wu is supported by the postdoctoral fellowship
from the Academia Sinica, Taiwan.

\end{document}